\input{aipcheck}

\documentclass[
    ,final            
  ]
  {aipproc}

\layoutstyle{8x11double}

\begin{document}

\title{Possible Detection of a Pair Instability Supernova in the Modern
Universe, and Implications for the First Stars}



\author{Nathan Smith}{
  address={Astronomy Department, University of California, Berkeley} }


\begin{abstract}

SN~2006gy radiated far more energy in visual light than any other
supernova so far, and potential explanations for its energy demands
have implications for galactic chemical evolution and the deaths of
the first stars. It remained bright for over 200 days, longer than any
normal supernova, and it radiated more than 10$^{51}$ ergs of luminous
energy at visual wavelengths. I argue that this Type IIn supernova was
probably the explosion of an extremely massive star like Eta Carinae
that retained its hydrogen envelope when it exploded, having suffered
relatively little mass loss during its lifetime.  That this
occurred at roughly Solar metallicity challenges current paradigms for
mass loss in massive-star evolution.  I explore a few potential
explanations for SN2006gy's power source, involving either
circumstellar interaction, or instead, the decay of $^{56}$Ni to
$^{56}$Co to $^{56}$Fe.  If SN~2006gy was powered by the conversion of
shock energy into light, then the conditions must be truly
extraordinary and traditional interaction models don't work.  If
SN~2006gy was powered by radioactive decay, then the uncomfortably
huge $^{56}$Ni mass requires that the star exploded as a pair
instability supernova. The mere possibility of this makes SN~2006gy
interesting, especially at this meeting, because it is the first good
candidate for a genuine pair instability supernova.


\end{abstract}

\maketitle


\section{Introduction}

For the purposes of this meeting, the main relevant point is to ask
whether or not SN~2006gy was really a pair instability supernova
(PISN).  I'll get into that later, but the short version is that we
are not really sure.  Its basic observed properties (high luminosity,
long duration, and slow expansion speed of a heavy H envelope) seem
consistent with the basic attributes of a PISN.  On the other hand,
alternative scenarios -- while somewhat problematic and perhaps
equally extraordinary -- are hard to rule out conclusively.
Nevertheless, it is worth discussing SN~2006gy in the context of
PISNe, because if it was one, it may have some far-reaching
implications for the pollution of interstellar matter and for learning
about the deaths of the first stars.  First, though, I'll just list
some basic observables of SN~2006gy taken from Smith et al.\ (2007)
and Ofek et al.\ (2007):

\begin{itemize}

\item The host galaxy was NGC~1260, which is a peculiar S0/Sa galaxy
with sufficient evidence for star formation to make the presence of
very massive stars plausible.  SN~2006gy was about 300 pc from the
galaxy's nucleus, and may have been near a spiral arm.  The distance
to NGC~1260 is 73.1 Mpc.

\item The average metallicity of the host galaxy is roughly Solar
(Z$\simeq$0.63 Z$_{\odot}$).  We do not have a very good estimate of
the metallicity at SN~2006gy's specific position in the galaxy.

\item The peak bolometric luminosity was at least $\sim$5$\times$10$^{10}$
L$_{\odot}$ ($M_R \simeq -22$), and the total radiated energy in
visible light during the first $\sim$200 days was at least
1.4$\times$10$^{51}$ ergs.  This is 10 to 100 times more luminous than
typical Type Ia and core collapse supernovae, respectively, and more
than a factor of 100 greater than the total radiated energy of most
SNe.

\item In addition to being bright at its peak, the light curve of
SN~2006gy evolved very slowly, taking more than 70 days to reach that
peak, and even longer to decline (Fig.\ 1).  This is unlike other
SNe.

\item Spectroscopically, SN~2006gy was a Type IIn supernova, meaning
that it had narrow lines of hydrogen.  This indicates that the star
retained much of its original H envelope until the time it died.

\item The H$\alpha$ profile is actually rather complicated (Fig.\ 2),
showing a very narrow P Cygni feature from slow-moving circumstellar
material at 130-260 km s$^{-1}$, plus a broader emission component of
a few thousand km s$^{-1}$ (still much narrower than most supernovae).
The speed of a few hundred km s$^{-1}$ indicated by the narrow
component is too fast for a red supergiant (RSG) wind, but is
consistent with the wind of a luminous blue variable (LBV) like $\eta$
Carinae.

\item The expansion speed indicated by broad P Cygni absorption in the
H$\alpha$ line was about 4,000 km s$^{-1}$, which did not change
perceptably in the first $\sim$200 days.  In other words, the SN blast
wave did not decelerate much during a time when the SN emitted more
than 10$^{51}$ ergs of light.  That will be important later.

\item SN~2006gy was detected in soft X-rays by {\it Chandra}, and was
spatially resolved from the nucleus of the galaxy, which appears to be
an active nucleus.  Analysis of the X-rays detected near the time of
peak visual light of SN~2006gy implies a mass-loss rate for the
progenitor's wind of no more than about 5$\times$10$^{-4}$ M$_{\odot}$
yr$^{-1}$.  The few X-rays that were detected (only 4! ...but a
10$\sigma$ detection) were all soft X-rays.

\end{itemize}



\begin{figure}
  \includegraphics[height=.27\textheight]{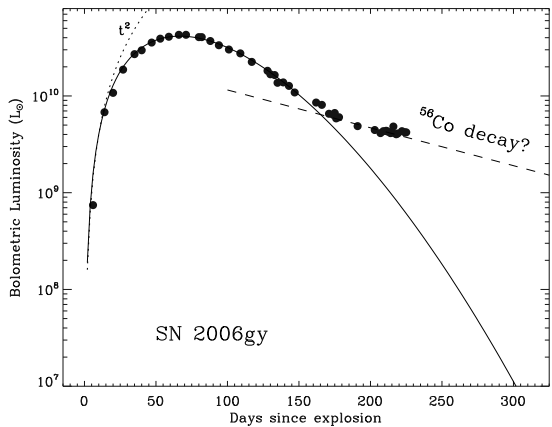}
  \caption{An approximation of the bolometric light curve of SN~2006gy
  adopting the unfiltered red magnitude with no bolometric correction
  (see Smith et al.\ 2007).  It may therefore be an underestimate of
  the true luminosity.  Although it uses the KAIT lightcurve from
  Smith et al.\ (2007), this figure is actually from Smith \& McCray
  (2008), using a simple thermal diffusion model to approximate the
  main part of the light curve, followed by a dashed line representave
  of radioactive decay from about 8 M$_{\odot}$ of Ni.}
\end{figure}

\begin{figure}
  \includegraphics[height=.31\textheight]{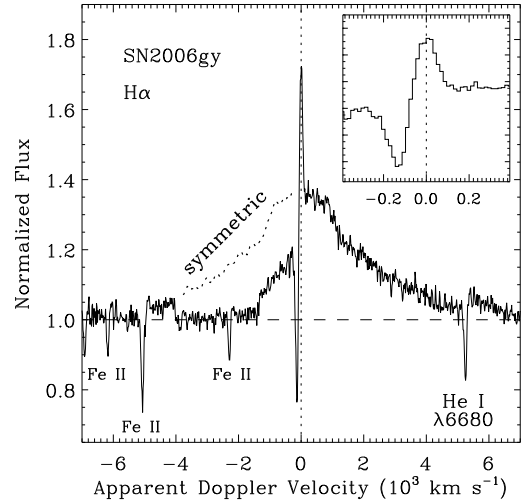} 
  \caption{The H$\alpha$ profile of SN~2006gy near the time its
  lightcurve peaked (from Smith et al.\ 2007).  The inset shows an
  expanded version of the narrow P Cygni feature from the CSM.  The
  dotted line is a reflection of the red side of the line, showing how
  a symmetric profile compares to the observed line shape.}
\end{figure}

\section{SN~2006\lowercase{gy} as a PISN?}

The idea of PISNe has been around for four decades (Barkat, Rakavy,
\& Sack 1967; Bond, Arnett, \& Carr 1984; Heger \& Woosley 2002), but
we have yet to observationally verify that such explosions exist.
While the case is still indefinite, SN~2006gy provides us with the
first good candidate that matches the basic expected properties of a
PISN.  Namely, as described by Smith et al.\ (2007), it fits
qualitative predictions for model lightcurves of PISNe (e.g.,
Scannapieco et al.\ 2005) in that it was extremely luminous, it had a
slow rise time and long duration, it had slow expansion speeds because
of the heavy envelope, and it was hydrogen rich.  While the light
curve of SN~2006gy does not precisely fit the predictions of
Scannapieco et al., we should remember that those models were for zero
metallicity stars with no mass loss, since that was the place PISNe
were generally expected to occur.  Direct application of those models
to massive stars in the modern Universe at roughly Solar metallicity
is probably unwise.  While the progenitor of SN~2006gy did retain much
of its original H enelope, it clearly has also suffered substantial
mass loss prior to explosion, evidenced by emission from dense
circumstellar material.  Including some mass loss (and changing the
mass of material through which radiation must diffuse) can drastically
alter the shape of the light curve that one calculates.  Some more
recent work on PISN models described at this meeting in talks by K.\
Nomoto, D.\ Kasen, and T.\ Young shows different predictions for the
light curves from different assumptions.

So far, the PISN model seems to be a viable explanation for SN~2006gy.
A critical test will be its future behavior; SN~2006gy is currently
behind the Sun and is unobservable.  If its late-time decline reflects
$^{56}$Co decay, then this can be proof positive that SN~2006gy was a
PISN.  We should remember, though, that there may be some trickery
involved.  For example, the visual/red light curve in Figure 1 is not
{\it really} a bolometric lightcurve.  As the SN expands and cools,
its bolometric flux may shift to the red and infrared, so a careful
assessment of the multiwavelength properties need to be considered
before we can say that it has faded faster than the radioactive decay
rate (in fact, this may be part of the explanation for the
leveling-off of the light curve at late times in Fig.\ 1).  Also,
late-time photometry will be tricky because of the background galaxy
light, and then there's always the possibility of obscuration from
dust formation. Thus, we should expect that sensitive IR observations
will be critical to the nature of SN~2006gy in the immediate future.
Another approach to evaluate our confidence in the viability of the
PISN is to examine possible alternative explanations, as described
next.

\section{Alternatives: Conversion of Shock Energy Into Light}

Efficient conversion of blast wave kinetic energy into visual light is
the only potentially-viable alternative to radioactive decay as a
power source for the tremendous luminosity and total radiated energy
of SN~2006gy.  This is the current best-bet explanation for the
luminosity and spectral properties of the class of Type IIn supernovae
(e.g., $ref$), which can be more luminous than normal Type II-P
supernovae, and show evidence for dense circumstellar material (CSM)
in their spectra.  The spectrum of SN~2006gy showed some clear
characteristics in common with the Type IIn class (Smith et al.\ 2007;
Ofek et al.\ 2007), including its relatively narrow H$\alpha$ line.
Aspects of the CSM interaction hypothesis for SN~2006gy have been
discussed in detail by Ofek et al.\ (2007), Smith et al.\ (2007),
Woosley et al.\ (2008), and Smith \& McCray (2008).  The main
consideration for the CSM-interaction hypothesis is that because of
the extraordinarily-high luminosity of SN~2006gy, the CSM must
be very dense and massive.  I'll return to this momentarily.

The basic idea of how an interaction model can generate the extra high
luminosity is that the SN blast wave expands out into a dense CSM.
The CSM is swept up into a dense post-shock cooling layer that emits
the observed visual continuum luminosity.  In that scenario, the high
luminosity phase can last until the shock runs past the densest CSM,
or until the shock runs out of energy (you might call these cases
``density bounded'' or ``energy bounded'', reminiscent of terms for
H~{\sc ii} regions).

Ofek et al.\ (2007) first suggested this type of CSM-interaction model
for SN~2006gy, discussed primarily in the framework of a Type Ia
supernova interacting into a dense H-rich CSM.  This is similar to
what had been proposed for the recent luminous SNe 2002ic and 2006gj
(Hamuy et al.\ 2003; Aldering et al.\ 2006).  Smith et al.\ (2007) and
Woosley et al.\ (2008) favored a somewhat different interpretation
that SN~2006gy had been a very massive star that created its dense CSM
when it suffered a giant outburst analogous to that of $\eta$~Carinae,
occurring in the decade immediately before the SN.  Observationally,
such an outburst would probably have appeared similar to LBVs.  The
cause of LBV-type outbursts is not known, but one potential
explanation is that it may have been triggered by the pulsational pair
instability of Heger \& Woosley (2002).  Even with an explanation for
how to create the dense CSM, however, the case of SN~2006gy presents
further constraints that complicate the normal Type IIn or
CSM-interaction interpretation..


Because of the extreme energy demands of SN~2006gy, there are a couple
things to keep in mind as we are considering a CSM-interaction model
that we don't usually need to worry about in fainter Type IIn
supernovae.  First, since the total radiated energy is more than about
10$^{51}$ ergs (Smith et al.\ 2007), the very efficient conversion of
kinetic energy to light would drain a canonical SN blast wave of
essentially {\it all} its available energy.  Thus, for a normal SN,
the expansion speed deduced from the spectrum should decelerate and
eventually slow to a crawl as the energy is released as light.  This
did not happen in SN~2006gy, which showed a relatively constant
expansion speed of about 4000 km s$^{-1}$ throughout the time when it
emitted its tremendous 10$^{51}$ ergs of visual light (Smith et al.\
2007).  This requires either that some other source powers the visual
light (like $^{56}$Co decay), or that the available reservoir of KE was
much more than 10$^{51}$ ergs.  Either case is exotic.

Second, even if we allow high efficiency in converting KE into light
and we allow for very energetic explosion, we still run into other
fundamental obstacles having to do with radiative transfer.  In order
to convert more than 10$^{51}$ ergs of KE into light, a great deal of
mass is required in the CSM -- roughly 10 M$_{\odot}$ or more (Smith
et al.\ 2007; Ofek et al.\ 2007) --- ejected in the decade or so just
before the SN.  The fundamental problem, as pointed out recently by
Smith \& McCray (2008), is that this amount of mass in the CSM would make
that same CSM very opaque, so that {\it the radiation could not escape
even if it could be generated!}  This makes it impossible for
conventional CSM-interaction models to explain SN~2006gy, because they
rely on the continuous generation of visual light as the blast wave
sweeps through the CSM.

Smith \& McCray (2008) suggest a possible way to circumnavigate this
paradox that may rescue the CSM-interactionhypothesis for SN~2006gy.
It may still draw its power from the conversion of KE into light, but
in a way that is different from the continous CSM interaction models
suggested by Ofek et al.\ (2007) and Woosley et al.\ (2007).  Namely,
Smith \& McCray proposed that the visual luminosity results from the
diffusion of shock-deposited thermal energy escaping from an extended
circumstellar envelope {\it after} it has already been overrun and
accelerated by the blast wave.  This is analogous to the Type II
models of SNe in extended red supergiant envelopes (Falk \& Arnett
1973), except that here the envelope is not bound to the star, having
been ejected in an LBV-like event a few years earlier.  The $\sim$10
M$_{\odot}$ LBV envelope was initially very opaque ($\tau\simeq$300)
at the time the SN occurred, and the delayed escape of radiation then
occurs as photons diffuse out of the shocked envelope as it expands
and thins.  See Smith \& McCray (2008) for further details.

Now, SN~2006gy did have a spectrum that resembled Type IIn supernovae,
as noted above, suggesting {\it ongoing} CSM-interaction as well.  How
can that be if the shock interaction with the shell was opaque?  Well,
after the blast wave breaks free of the opaque shell (which is when
the visual light curve begins to rise) it will continue to expand into
whatever CSM happens to reside there -- which we might expect to be a
normal (i.e. non-outburst) stellar wind.  In fact, the mass-loss rate
derived from the usual interaction diagnostics like H$\alpha$ and the
soft X-rays detected by {\it Chandra} indicated a fairly modest
mass-loss rate that appeared entirely consistent with a normal stellar
wind from a luminous blue supergiant star (Smith et al.\ (2007).  That
lower mass-loss rate would be nowhere near enough to account for the
continuum luminosity of SN~2006gy, but it could give rise to the
characteristic Type IIn spectrum.

The continuum luminosity needs a different source.  Smith et al.\
(2007) noted several additional difficulties with a conventional CSM
interaction model that could be explained by a PISN if the visual
luminosity arises in the ejecta and not the ongoing interaction region
In the photon diffusion model of Smith \& McCray (2008), the emitting
geometry is similar to the situation where the bulk of the continuum
luminosity arises in central stellar ejecta and not the ongoing CSM
interaction region, so this model may be able to satisfy the
obserational constraints as well as the PISN hypothesis.  Either one
is still viable at this point, but late-time observations may soon be
able to give us an answer.



\section{Mass loss in the evolution of massive stars}

Whatever interpretation correctly explains the high luminosity of
SN~2006gy, it is clear that the extreme energy demands require the
death of a very mass star that may have started its life near the
upper mass limit for stars.  The facts that this star died with much
of its H envelope intact and that it apparently had a mysterious
outburst just before its final death as a SN are not understood in our
current understanding of stellar evolution.  A very massive star dying
without shedding much of its mass also smells a little bit like the
first stars, but this one happened at roughly Solar metallicity.

How can this be?  Recent years have seen a revision in our thoughts
about mass loss in the evolution of massive stars.  One important clue
is that several lines of evidence point to the fact that winds of hot
stars may be highly clumped.  This, in turn, means that mass-loss
rates derived from density-squared diagnostics like H$\alpha$ and
radio continuum emission have severely overestimated mass loss rates.
This conclusion derived from years of work done by many groups; a
summary and relevant references can be found in Smith \& Owocki
(2006).  The main conclusion (Smith \& Owocki 2006) is that
metallicity-dependent line-driven winds of O stars do not shed very
much mass in a star's lifetime, and therefore cannot account for the
stripping of H envelopes that harkens the formation of Wolf-Rayet (WR)
stars.

Nevertheless, WR stars do exist so some massive stars must be able to
shed their H envelopes by some other means.  Smith \& Owocki pointed
out that the best candidates for the mass-loss mechanism that may make
up the difference are the continuum-driven eruptions of LBVs, where a
star can shed 10 M$_{\odot}$ or more in a single event that lasts less
than a decade.  These event are observed, but we still do not
understand what triggers them.

On the other hand, the existence of Type IIn SNe like SN~2006gy and
others indicates that some massive stars evolve differently -- they
{\it do not} fully shed their H envelopes, even through LBV eruptions,
and die before they can become a WR star.  In that case, a major goal
in massive star evolution will be to determine how important these
episodic outbursts are at different values of metallicity, initial
mass, initial rotation rates, etc., and re-evaluating the fates of
massive stars as these quantities vary.

\section{Implications for the first stars}

A subtle implication of SN~2006gy is that in some respects, and in
some special cases, the evolution and deaths of very massive stars in
the modern Universe might not be so terribly different from the first
stars after all -- at least in terms of their mass loss properties.
If SN~2006gy was a PISN, it obviously has potential to inform our
views about the deaths of the first stars and our interpretations of
the role of PISNe in galactic chemical evolution.  Even if SN~2006gy
turns out not to have been a PISN, however, the fact remains that it
reached the end of its life and exploded without losing much of its
initial mass.  This may suggest that studying massive stars in the
local Universe may still tell us some things about the evolution of
the first stars.  Namely, with much lower mass-loss rates, the
mass-loss evolution and angular momentum evolution of modern stars
may, in some cases, be similar to early stars.

By the same token, very massive stars in the local Universe may then
offer some sobering words of caution about our understanding of the
first stars.  For example, in recent years it has turned out that many
of our assumptions about mass loss and the evolution of massive stars
at Solar metallicity has turned out to be wrong.  It seems to be the
case that the most massive stars do not shed most of their mass in
metallicity-dependent line-driven winds, and can instead shed much
more mass in continuum-driven outbursts like the outbursts of LBVs
(see Smith \& Owocki 2006).  This is for stars in the local Universe
where we have been collecting observational data for many decades.
What does this imply about our understanding of the first stars, for
which it will be some time before we have any direct observational
tests?

Finally, the simple fact tha SN~2006gy was so bright suggests that
there is some fraction of SNe -- perhaps a small fraction -- that can
potentially be detected to much greater distances than conventional
SNe.  If SN~2006gy was indeed a PISN, then this obviously bodes well
for our hopes of directly detecting the first SNe with future
instruments like JWST.  Again, though, even if SN~2006y was not a true
PISN, there may be hope of detecting other SNe like it far back in
time.  If it wasn't a PISN, then the reason SN~2006gy was so bright
was because it suffered a giant LBV-like mass ejection just a few
years before it died.  Smith \& Owocki (2006) have argued that these
events are driven by a mechanism that appears to be independent of
metallicity.  Since this metallicity-independent mechanism aparently
dominates the mass loss of very massive stars in the local Universe,
we should not rule out the possibility that it may also be important
at very low metallicity in the early Universe.  If so, LBV-like blasts
followed by SN could make similarly luminous events, and could be
detected in the early Universe as well.

OK -- just one more final thought.  So far, the only empirical data
that has been mentioned in connection with the first stars and PISNe
is the abundance patterns in very very very low-metallicity stars
thought to have been born in the early Universe, and there have been
many illuminating talks on that subject at this meeting.  Is there an
analogous way to search for evidence of any signature of modern PISNe,
such as in meteoritic abundance patterns or abundances of young stars
or protostellar disks at the edges of very massive giant H~{\sc ii}
regions where a very massive star might have already exploded?
Perhaps this may be an avenue to rule out the idea that PISNe occur
locally.


\begin{theacknowledgments}

This contribution summarizes some of my own thoughts, but it follows
directly from the research on SN~2006gy done with my collaborators:
W.\ Li, R.J.\ Foley, J.C.\ Wheeler, D.\ Pooley, R.\ Chornock, A.V.\
Filippenko, J.M.\ Silverman, R.\ Quimby, J.S.\ Bloom, C.\ Hansen, and
R. McCray.  SN research under A.V.\ Filippenko at UC Berkeley is
supported by NSF grant AST-0607485 and the TABASGO Foundation, and I
thank the conference organizers for partial financial support.

\end{theacknowledgments}


\end{document}